\documentclass[aps,showpacs,prb]{revtex4}
\usepackage{graphicx}
 
\newcommand{\be}{\begin{eqnarray}}
\newcommand{\ee}{\end{eqnarray}}
\newcommand{\Tr}{\mbox{Tr}\,}
\newcommand{\bms}{\mbox{\boldmath{$s$}}}

\begin{document}
\title{Dynamical correlations in the Sherrington-Kirkpatrick model 
in a transverse field}
\author{Kazutaka Takahashi}
\affiliation{Department of Physics, Tokyo Institute of Technology,  
 Tokyo 152--8551, Japan}
\author{Koujin Takeda}
\affiliation{Department of Computational Intelligence and Systems Science, 
 Tokyo Institute of Technology, Yokohama 226--8502, Japan}
\date{\today}

\begin{abstract}
 We calculate the real-time-correlation function 
 of the Sherrington-Kirkpatrick spin-glass model in a transverse field.
 Using a careful analysis of the perturbative expansion of
 the functional-integral representation,
 we derive the asymptotic form of the correlation function.
 In contrast to the previous works,
 we find for large transverse field 
 a power-law decay of the correlation 
 with time $t$ as $t^{-3/2}$ at zero temperature 
 and $t^{-2}$ at infinite temperature.
 At the small field region, 
 we also find a significant change in the correlation which comes 
 from the structural difference between two paramagnetic phases.
\end{abstract}
\pacs{
75.10.Nr, 
75.40.Gb, 
05.30.-d 
}
\maketitle

\section{Introduction}

 Among disordered systems, 
 spin glasses \cite{MPV} are of great interest
 and have been investigated over decades
 because of the simplicity of the models, 
 the rich resulting properties, and a large amount of applications to
 a variety of problems such as neural networks and 
 information processing.~\cite{Nishimori}
 The Sherrington-Kirkpatrick (SK) model \cite{SK} is 
 one of the objectives which has been intensively investigated 
 and is also known as the most successful example which yields
 a large number of notable consequences by mean-field approach.
  
 Quantum random magnets are also the
 objectives of challenge and of significance due to
 their relevance to the physics of real disordered 
 magnetic compounds at low temperature.
 However there is an obstacle. 
 Combination of randomness and quantum
 fluctuation makes the analytical approach complicated.
 Physically it is expected that spin-glass state may
 be disturbed by quantum fluctuations at low temperature and
 the system exhibits quantum phase transitions,
 which was discussed in several former works.~\cite{BM,CDS,Sachdev}
 
 In this paper we single out the SK model in a transverse field among
 a variety of quantum random magnet models 
 because this is the simplest quantum model exhibiting 
 the spin-glass phase transition and is believed to have 
 direct relationship with the physics of a real disordered compound 
 ${\rm LiHo}_{x}{\rm Y}_{1-x}{\rm F}_4$.
 We concentrate on clarifying the role of 
 the quantum fluctuations for spin-glass systems similar to former works.
 We use the path-integral formalism
 for dealing with the quantum fluctuation.  
 The effect of quantum fluctuation can be incorporated
 as (imaginary) time-dependent order parameters. 
 In Ref.~\onlinecite{KT}, one of the authors 
 proposed a method for treating such fluctuation effect,
 where the time dependence of the order parameters is integrated out 
 for deriving a renormalized effective free energy expressed 
 in terms of classically defined time-independent order parameters. 
 This approach is sufficient for extracting static properties 
 of the system, e.g., the phase diagram or time-independent quantities, 
 which can be assessed by the resulting effective free energy.
 However in this paper we look closely at the time dependence of 
 the order parameter to investigate the dynamical properties of the system.
 Therefore we resort to another approach to assess time-dependent quantities. 

 We analyze the local dynamical correlation function 
 at an arbitrary temperature $T=1/\beta$, which is defined as 
\be
 \chi(t) 
 = \left[\frac{1}{Z}
 \Tr e^{-\beta\hat{H}}e^{i\hat{H}t}\sigma^z_i
 e^{-i\hat{H}t}\sigma^z_i\right], \label{chit}
\ee
 where $\sigma^z_i$ is the $z$ component Pauli-spin operator on site $i$ 
 and $Z = \Tr\exp(-\beta \hat{H})$ is the partition function
 with a Hamiltonian $\hat{H}$.
 In the analysis of quantum systems, the calculation of 
 the dynamical correlation functions is one of the main topics 
 and a variety of analytical techniques 
 have been developed for it.~\cite{Sachdev,GKKR}
 As stated above, we consider 
 the SK model in a transverse field $\Gamma$, 
\be
 \hat{H} = -\frac{1}{2}\sum_{i,j=1}^{N}J_{ij}\sigma^z_i\sigma^z_j
 -\Gamma\sum_{i=1}^N\sigma^x_i, \label{GSK}
\ee
 where the averaging of the spin coupling $J_{ij}$,
 denoted by the square brackets in Eq.(\ref{chit}), 
 is taken with the Gaussian distribution.
 
 With regard to the behavior of dynamical correlation function 
 there are many preceding analytical \cite{MH,YSR,RG,GR} and 
 numerical \cite{ARi,AR} works, where Eq.(\ref{chit}) 
 with imaginary time $t=-i\tau$ is analyzed.
 Their analytical results show that 
 the correlation decays asymptotically as power law of $\tau^{-2}$ 
 at the zero-temperature critical point.
 Furthermore the authors of Refs.~\onlinecite{RG} and 
 \onlinecite{GR} pointed out the relevance of $\tau^{-2}$ decay 
 to the same behavior of time-correlation 
 function in the single-impurity Kondo model \cite{Kondo} 
 by making use of the mapping between two models,
 which seems to enforce the validity of their result.
 However, the numerical calculation in Ref.~\onlinecite{ARi}
 indicates a somewhat smaller value for the power-law index
 and some argument is required for the discrepancy.
 All of the former analytical results are based on perturbative
 expansion, and in this paper we reexamine this calculation by 
 using a careful analysis of the expansion.
 We arrive at a different conclusion for the value of the power index,
 which stems from how to deal with multipoint correlation function 
 appearing in the form of expansion.

 The paper is organized as follows.
 In Sec.~\ref{zero}, we calculate the imaginary-time-correlation 
 function at zero temperature by using the perturbative expansion.
 By referring to the diagrammatic expansion method 
 utilized for other quantum systems,
 we calculate the asymptotic form of the correlation function
 and make the comparison of it with the known results.
 In addition the result is confirmed by the numerical calculation.
 In Sec.~\ref{finite} we move on to the correlation at finite temperature.
 We investigate it numerically and
 give some considerations to the result, which suggest different
 asymptotic behaviors of the correlation at high-temperature region
 from the case of zero temperature.
 Section \ref{conc} is devoted to conclusions.

\section{Correlations at zero temperature}
\label{zero}

\subsection{Imaginary-time formalism}

 The real-time correlation (\ref{chit}) is obtained by 
 the analytic continuation of the imaginary-time-correlation function, 
\be
 \chi(\tau)
 = \left[\frac{1}{Z}
 \Tr e^{-(\beta-\tau)\hat{H}}
 \sigma^z_ie^{-\tau\hat{H}}\sigma^z_i
 \right]. \label{chitau}
\ee
 The time-independent static part of $\chi(\tau)$,
 $\chi=\int_0^\beta d\tau\chi(\tau)/\beta$, was studied 
 in detail in Ref.~\onlinecite{KT} 
 as a measure of quantum fluctuations.
 In order to deal with quantum operators analytically, 
 we use the imaginary-time path-integral representation.
 Following the standard technique 
 we introduce replica to perform the ensemble average. 
 The correlation function is expressed with 
 the path integral of normalized vector, 
\be
 \chi(\tau)
 &=& \lim_{n\to 0}\left[\Tr 
 \exp\left(-\beta \sum_{\alpha=1}^n\hat{H}^{(\alpha)}\right)
  \sigma^{(1)}_{zi}(\tau)\sigma^{(1)}_{zi}(0)
 \right] \\
 &=&
 \lim_{n\to 0}\left[ \int{\cal D}{\bms} \,
 s^{(1)}_{zi}(\tau)s^{(1)}_{zi}(0)
 \exp\left(
 \sum_{\alpha=1}^n\int_0^\beta d\tau' \left\{
 i\Phi[\bms^{(\alpha)}(\tau')]-H[\bms^{(\alpha)}(\tau')]
 \right\}\right)
 \right],
\ee
 where the superscript $\alpha$ is the replica index 
 running from 1 to $n$, and the limit $n\to 0$ is taken afterward.
 $\bms$ is the unit vector on the Bloch sphere and 
 $\Phi$ is the Berry phase term.
 Then we put the Hamiltonian of the transverse SK model (\ref{GSK})
 into the above formula and take the Gaussian average 
 with respect to the interaction $J_{ij}$.
 After performing the Hubbard-Stratonovich transformation, 
 we have the path integral form of one-body Hamiltonian,
\be
 \chi(\tau) = \left<
 s^{(1)}_{z}(\tau)s^{(1)}_{z}(0)
 \exp\left\{
 \frac{J^2}{2}\sum_{\alpha=1}^n\int_0^\beta d\tau_1\int_0^\beta d\tau_2\,
 s_{z}^{(\alpha)}(\tau_1)\chi(\tau_1-\tau_2)s_{z}^{(\alpha)}(\tau_2)
 \right\}
 \right>_\Gamma,
 \label{sceq}
\ee
 where the angular bracket denotes the path integral over
 $\bms^{(\alpha)}(\tau)$ as
\be
 \langle \cdots\rangle_\Gamma = 
 \lim_{n\to 0}\int{\cal D}\bms \left(\cdots\right)
 \exp\left(
 \sum_{\alpha=1}^n\int_0^\beta d\tau' \left\{
 i\Phi[\bms^{(\alpha)}(\tau')]+\Gamma s_{x}^{(\alpha)}(\tau')
 \right\}\right).
\ee
 The Hubbard-Stratonovich field $\chi(\tau_1-\tau_2)\sim \langle 
 s_z^{(\alpha)}(\tau_1)s_z^{(\alpha)}(\tau_2)\rangle$ 
 is nothing but the time-correlation function (\ref{chitau})
 and is taken to be free from the replica index.
 Here we neglect the time correlation between different replicas,
 or in other words time-dependent spin-glass parameter
 $q^{\alpha\alpha'}(\tau_1,\tau_2)\sim \langle 
 s_z^{(\alpha)}(\tau_1)s_z^{(\alpha')}(\tau_2)\rangle$
 $(\alpha\ne\alpha')$ is taken to be zero,
 which means we are in paramagnetic phase.
 Our goal is to solve this self-consistent equation for $\chi(\tau)$.

\subsection{Self-consistent equation in linear order}
\label{linear}

\begin{figure}[tb]
\begin{center}
\includegraphics[width=0.5\columnwidth]{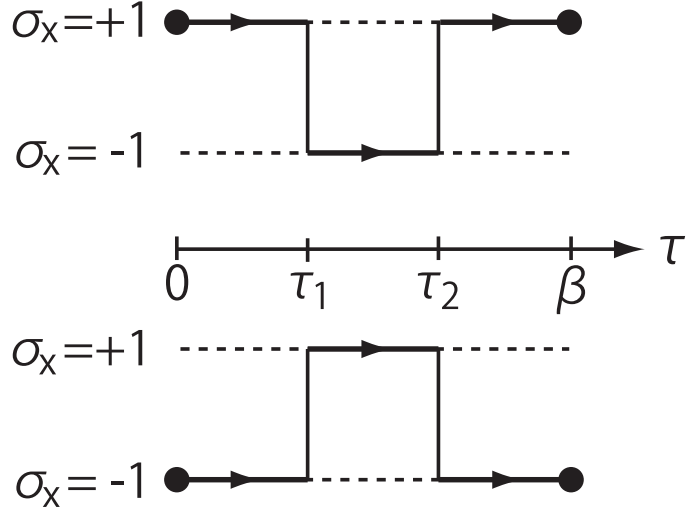}
\caption{The pictorial representation of the two-point correlation function 
 $D_\Gamma(\tau_2-\tau_1)=\langle s_z(\tau_2)s_z(\tau_1)\rangle_\Gamma$
 with $\tau_2>\tau_1$.}
\label{d}
\end{center}
\end{figure}

 We expand Eq.(\ref{sceq}) as a power series of $J^2$ 
 in order to perform the spin integration afterward.
 Up to the first order, we find 
 the self-consistent linear equation for $\chi(\tau)$, 
\be
 \chi(\tau) = D_\Gamma(\tau) +\frac{J^2}{2}
 \int_0^\beta d\tau_1 \int_0^\beta d\tau_2\left\{
 D_\Gamma(\tau,\tau_1,\tau_2,0)-D_\Gamma(\tau)D_\Gamma(|\tau_1-\tau_2|)
 \right\}\chi(\tau_1-\tau_2), \label{chitaueq1}
\ee
 where the two-point correlation function 
 $D_\Gamma(|\tau_1-\tau_2|) 
 = \langle s_z^{(1)}(\tau_1)s_z^{(1)}(\tau_2)\rangle_\Gamma$
 is given by
\be
 D_\Gamma(|\tau_1-\tau_2|) 
 = \frac{e^{\beta\Gamma-2\Gamma|\tau_1-\tau_2|}
 +e^{-\beta\Gamma+2\Gamma|\tau_1-\tau_2|}}
 {e^{\beta\Gamma}+e^{-\beta\Gamma}}.
\ee
 We can easily find
 how two $s_{z}$'s contribute to this propagator
 by drawing pictorial spin-flip representation as in Fig.~\ref{d}.
 The spin pointing along the $x$ direction 
 is flipped two times by the spin operator $\sigma_z$
 while propagating from $\tau=0$ to $\tau=\beta$.
 It is necessary to go back to the original state
 due to the boundary condition $\bms(0)=\bms(\beta)$.~\cite{KT}
 In the same way, for the four-point function
 $D_\Gamma(\tau_1,\tau_2,\tau_3,\tau_4)
 =\langle s_z^{(1)}(\tau_1)s_z^{(1)}(\tau_2)s_z^{(1)}(\tau_3)
 s_z^{(1)}(\tau_4)\rangle_\Gamma$, 
 the spin is flipped four times and we obtain 
\be
 D_\Gamma(\tau_1,\tau_2,\tau_3,\tau_4) = 
 D_\Gamma(\tau_1-\tau_2+\tau_3-\tau_4),
 \label{4pt}
\ee
 where we assume $\tau_1>\tau_2>\tau_3>\tau_4$.
 As we see from Fig.~\ref{d}, 
 it is important to take notice of the time ordering
 of the spin operators.

 The integral equation (\ref{chitaueq1}) can be solved 
 by the differential equation for $\chi(\tau)$, 
\be
 \partial_\tau^4\chi(\tau)-8\Gamma^2\partial_\tau^2\chi(\tau)
 +16(\Gamma^4-\Gamma^2 J^2)\chi(\tau) = 0,
\ee
 which is obtained by taking derivatives of 
 Eq.(\ref{chitaueq1}) several times and 
 constructing the closed-form equation in terms of $\chi(\tau)$. 
 With the boundary condition it can be solved as
\be
 \chi(\tau) = 
 \frac{1}{2}\left[
 D_{\Gamma_+}(\tau)+D_{\Gamma_-}(\tau)
 \right], \label{chitau1}
\ee
 where $\Gamma_\pm=\sqrt{\Gamma^2\pm\Gamma J}$.
 At the zero-temperature limit, this function goes to 
\be
 \chi(\tau) \to \frac{1}{2}\left(
 e^{-2\Gamma_+\tau}+e^{-2\Gamma_-\tau}\right), \label{chitau10}
\ee
 where we assume $\tau<\beta/2$ in taking the limit $\beta\to\infty$.
 This result shows that, due to disorder,  
 the energy level splits into two as 
 $\Gamma\to\sqrt{\Gamma^2\pm\Gamma J}$.
 Proceeding further to the higher order correlations, 
 we expect energy levels split into many 
 and finally form a broad distribution, 
 as we see in the following.

 However, we must note that 
 this interpretation is valid only when $\Gamma>J$.
 Solution (\ref{chitau1}) changes its behavior
 in the case of smaller transverse fields,
 which is attributed to the phase transition between 
 the paramagnetic and the spin-glass phases.~\cite{MH}
 When $\Gamma<J$, the time-independent part 
 $\chi=\int_0^\beta d\tau\chi(\tau)/\beta$ is nonzero 
 even at zero temperature and 
 nonperturbative analysis is required for the zero mode.~\cite{KT}
 This sharp change in behavior could be 
 an artifact of the perturbative expansion.
 Although this change survives even if we take the higher-order 
 correlations into account as we show in the following, it is not 
 clear whether it corresponds to the phase-transition point or not.
 For example, in the random energy model \cite{REM}
 the corresponding phase transition is of first order and 
 it is not possible to determine the transition point 
 from perturbative calculation 
 (see Sec.~\ref{para} for detailed calculations). 
 In the present analytical calculation, 
 we assume that $\Gamma$ is large enough 
 such that the zero mode gives no contribution. 
 Then the correlation asymptotically decays to zero in time 
 and the perturbative expansion is justified.
 The behavior at small $\Gamma$ is analyzed numerically 
 in Sec.~\ref{num}.

\subsection{Asymptotic behavior at zero temperature}

 We take higher-order correlations into account in the following.
 In general $\chi(\tau)$ including all of such corrections
 is not tractable, but in the case of the zero-temperature limit
 the evaluation turns out to be considerably simpler.
 The integral in Eq.(\ref{chitaueq1}) 
 consists of contributions from  
 (i) $\tau_{1,2}<\tau$, 
 (ii) $\tau_{1,2}>\tau$, 
 and (iii) $\tau_1<\tau<\tau_2$ or $\tau_2<\tau<\tau_1$. 
 As we found above, the first-order result (\ref{chitau10})
 is short range and decays exponentially with respect to $\tau$.
 Then we expect that the asymptotic form of $\chi(\tau)$ 
 is dominated by contributions from (i) and (ii) 
 since the constraint in (iii), $\tau_1\sim\tau_2\sim\tau$, 
 is harder than that in (i) and (ii), $\tau_1\sim\tau_2$,
 in performing double integral, 
 which makes the contribution from (iii) negligible.
 At the limit $T=0$, 
 the contribution from (i) is written as 
\be
 J^2\int_0^\tau d\tau_1\int_0^{\tau_1} d\tau_2 \left[
 g(\tau-\tau_1)\chi(\tau_1-\tau_2)g(\tau_2)
 -g(\tau-\tau_2)\chi(\tau_1-\tau_2)g(\tau_1)
 \right], \label{i}
\ee
 where $g(\tau)=\exp(-2\Gamma\tau)$.
 Similarly the contribution from (ii) is given by 
\be
 J^2\int_\tau^\beta d\tau_2\int_{\tau_2}^{\beta} d\tau_1 \left[
 g(\beta-\tau_1)\chi(\tau_1-\tau_2)g(\tau_2-\tau)
 -g(\beta-\tau_2)\chi(\tau_1-\tau_2)g(\tau_1-\tau)
 \right]. \label{ii}
\ee
 Using the relation $\chi(\tau)=\chi(\beta-\tau)$, 
 we find that Eq.(\ref{ii}) is given by Eq.(\ref{i}) 
 with the replacement $\tau$ by $\beta-\tau$.
 Comparing each term in above two equations
 we arrive at the conclusion that the main contribution 
 comes from the first term in Eq.(\ref{i}).
 We have an approximated form now, 
\be
 \chi(\tau) \sim g(\tau)
 +J^2\int_0^\tau d\tau_1 \int_0^{\tau_1}d\tau_2
 g(\tau-\tau_1)\chi(\tau_1-\tau_2)g(\tau_2).
\ee
 This equation is solved as
\be
 \chi(\tau) = \frac{1}{2}\left[
 e^{-2(\Gamma+J/2)\tau}+e^{-2(\Gamma-J/2)\tau}\right],
\ee
 which is a good approximation for Eq.(\ref{chitau10})
 when $\Gamma\gg J$.

 Higher-order contributions can be incorporated in the same way.
 We find
\be
 \chi = g+ J^2 g*\chi*g +J^4 g*\chi*g*\chi*g +\cdots 
 = \frac{1}{1-J^2 g*\chi}*g, \label{series}
\ee
 where the asterisk $*$ represents a time integral in convolution form, 
\be
 f*g(\tau) = \int^\tau_0 d\tau' f(\tau-\tau')g(\tau').
 \label{conv}
\ee
 This function product satisfies commutativity $f*g=g*f$
 and can be treated similar to ordinary product.
 Solving Eq.(\ref{series}) we obtain 
\be
 \chi = \frac{1}{2J^2 g}\left(
 1-\sqrt{1-4J^2 g^2}\right) 
 = \sum_{n=0}^\infty \frac{(2n-1)!!}{(2n+2)!!}
 2^{2n+1}J^{2n}g^{2n+1}, \label{seriesa}
\ee
 where we omitted the asterisk $*$.
 Using the $n$th convolution product of $g$ given by 
\be
 g^n(\tau) = \frac{\tau^{n-1}}{(n-1)!}e^{-2\Gamma\tau},
\ee
 we obtain the result 
 for the imaginary correlation function at zero temperature, 
\be
 \chi(\tau) = 
 \frac{1}{J\tau}I_1(2J\tau)e^{-2\Gamma\tau},
\ee
 where $I_{1}$ is the modified Bessel function of order 1.
 This function has an asymptotic form of exponential decay, 
\be
 \chi(\tau) \sim 
 \frac{1}{\sqrt{4\pi}(J\tau)^{3/2}}e^{-2(\Gamma-J)\tau},
 \label{chitauasy}
\ee
 at $\Gamma>J$, 
 which changes to a power decay $\tau^{-3/2}$ at $\Gamma=J$.

 The analytic continuation to the real time is obtained as  
\be
 \chi(t)= 
 \frac{1}{Jt}J_1(2Jt)e^{-2i\Gamma t},
 \label{chita}
\ee
 where $J_1$ is the Bessel function of order 1.
 This shows $t^{-3/2}$ decay for arbitrary $\Gamma$ as
\be
 \chi(t) \sim 
 \frac{1}{\sqrt{\pi}(Jt)^{3/2}}e^{-2i\Gamma t}
 \cos\left(2Jt-\frac{3\pi}{4}\right).
 \label{chitasy}
\ee
 In the Fourier-transformed space, 
 the spectrum forms a semicircle with the center $\omega=2\Gamma$ as
\be
 \chi(\omega) 
 = \frac{\sqrt{4J^2-(\omega-2\Gamma)^2}}{J^2}
 \theta(2J-|\omega-2\Gamma|). 
 \label{chiwa}
\ee
 It is worth stressing here that the real-time correlation 
 always shows a power-law asymptotic decay (\ref{chitasy}), 
 while the imaginary-time correlation (\ref{chitauasy}) exhibits 
 a power-law behavior only at $\Gamma=J$.
 However, as we explained in Sec.~\ref{linear}, we must be careful 
 about identifying the point $\Gamma=J$ as a critical one.
 Actually, it is known by preceding works \cite{KT,MH,AR} 
 that the value of $\Gamma$ at the critical point 
 should be a larger one $\Gamma\sim 1.5J$.
 In order to observe the power-law decay numerically 
 in imaginary-time framework, we must identify the critical point, 
 which turns out to be a troublesome task.~\cite{ARi} 
 On the other hand, in the real-time framework 
 the power-law behavior can be observed 
 not only in the critical point but also 
 in a broad range of large $\Gamma$  
 where our perturbative calculation is justified.
 Therefore, we can check the universal power-law index of 3/2
 in the paramagnetic phase at zero temperature 
 by observing the semicircle form of the spectrum.

\subsection{Comparison with previous results}

 Our result shows that the correlation decays in power as $t^{-3/2}$.
 On the other hand, the authors of Ref.~\onlinecite{MH} 
 have the conclusion of $t^{-2}$ decay via a rather general argument.
 Their argument is based on a perturbative expansion 
 as we did in the present paper.
 Actually we obtained the same form (\ref{seriesa}) as theirs, 
 though there is a minor but crucial difference. 
 The difference originates from the evaluation of
 the $n$-point correlations with $n\ge 4$.
 For example, let us recall that we have the four-point correlation
 $D_\Gamma(\tau_1,\tau_2,\tau_3,\tau_4) 
 =\langle s_z^{(1)}(\tau_1)s_z^{(1)}(\tau_2)s_z^{(1)}(\tau_3)
 s_z^{(1)}(\tau_4)\rangle_\Gamma$
 obtained in Eq.(\ref{4pt}). 
 In the previous works,~\cite{MH,YSR,RG,GR} the Wick theorem, 
\be
 D_\Gamma(\tau_1,\tau_2,\tau_3,\tau_4) \to
 D_\Gamma(|\tau_1-\tau_2|)D_\Gamma(|\tau_3-\tau_4|)
 +D_\Gamma(|\tau_1-\tau_3|)D_\Gamma(|\tau_2-\tau_4|)
 +D_\Gamma(|\tau_1-\tau_4|)D_\Gamma(|\tau_2-\tau_3|),
 \label{fact}
\ee
 was applied to the evaluation of this four-point correlation.
 However, this expression is not compatible with the exact one (\ref{4pt}).
 It is evident that the integration of the spin variable is not 
 a Gaussian one, and accordingly Eq.(\ref{fact}) is not justified.
 In order to show that the use of the Wick theorem leads to a wrong result,
 we examine the self-consistent equation 
 in linear order (\ref{chitaueq1}).
 In our approach we have the solution as Eq.(\ref{chitau1}),
 and we should recall that the form of 
 four-point correlation in Eq.({\ref{4pt}}) was used there.
 On the other hand, let us suppose that factorization (\ref{fact}) 
 by two-point correlation is correct. 
 Then we can perform the Fourier transformation of $\chi(\tau)$ to find
\be
 \chi(\omega_n=2\pi n/\beta)
 = D_\Gamma(\omega_n)+\beta^2J^2D_\Gamma^2(\omega_n)\chi(\omega_n) 
 = \frac{D_\Gamma(\omega_n)}{1-\beta^2J^2 D_\Gamma^2(\omega_n)},
\ee
 where $n$ is an integer and 
\be
 D_\Gamma(\omega_n) = \frac{\tanh\beta\Gamma}{\beta\Gamma}
 \frac{(\beta\Gamma)^2}{(\beta\Gamma)^2+(\pi n)^2}. 
\ee
 Performing the Fourier transformation again for going back to 
 the imaginary-time representation, we obtain
\be
 \chi(\tau) = \frac{1}{2}
 \left[
 \frac{\Gamma\tanh\beta\Gamma}{\tilde{\Gamma}_+\tanh\beta\tilde{\Gamma}_+}
 D_{\tilde{\Gamma}_+}(\tau)
 +\frac{\Gamma\tanh\beta\Gamma}{\tilde{\Gamma}_-\tanh\beta\tilde{\Gamma}_-}
 D_{\tilde{\Gamma}_-}(\tau)
 \right], \label{chitau1f}
\ee
 where $\tilde{\Gamma}_\pm =\sqrt{\Gamma^2\pm\Gamma J\tanh\beta\Gamma}$.
 This does not coincide with the exact result (\ref{chitau1}), which
 indicates that factorization (\ref{fact}) is not valid.
 In addition, Eq.(\ref{chitau1f}) does not satisfy
 the normalization condition $\chi(\tau=0)=1$.
 The authors of Refs.~\onlinecite{MH,YSR,RG,GR} introduced 
 a Lagrange multiplier to cure this defect.
 However, we can show that the correct result is not
 reproduced even if the Lagrange multiplier is introduced.
 Consequently we conclude that the factorization is not
 a correct procedure in evaluating the time-correlation function.

 Suppose again that we can apply the Wick theorem to the analysis here, 
 then the integration range of the convolution in Eq.(\ref{conv}) 
 is bounded by $\beta$, not by $\tau$,
 which means that the time ordering is not respected.
 Hence we may consider Fourier transformation of Eq.(\ref{seriesa}) as
\be
 \chi(\omega_n)= \frac{1}{2\beta^2 J^2 D_\Gamma(\omega_n)}
 \left(1-\sqrt{1-4\beta^2 J^2 D_\Gamma^2(\omega_n)}\right),
\ee
 which was found in Ref.~\onlinecite{GR}.
 This has the same form as Eq.(\ref{seriesa}), 
 but the time ordering is not treated properly  
 and a different result from Eq.(\ref{chiwa}) is obtained.
 
 The imaginary-time correlation was investigated numerically in 
 Ref.~\onlinecite{ARi} using the Monte Carlo method.
 The asymptotic form was obtained as $t^{-\alpha}$ 
 with $\alpha\sim 1.2$, which was considerably smaller than 
 the previously predicted value $\alpha=2$.
 Although the authors of Ref.~\onlinecite{ARi} 
 interpreted this value as the one which is close to $\alpha=1$,
 we consider this supports our result $\alpha=1.5$, 
 rather than the ones by former works.

\subsection{Numerical calculation}
\label{num}

\begin{figure}[tb]
\begin{center}
\includegraphics[width=0.5\columnwidth]{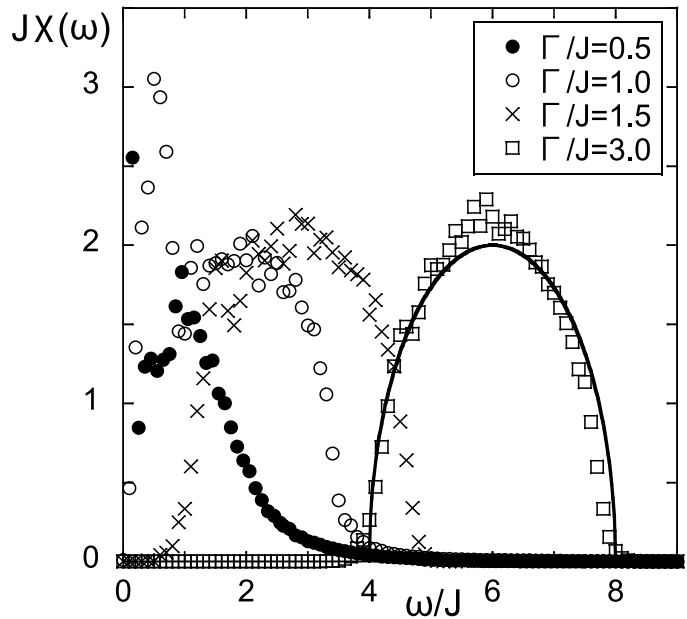}
\caption{The correlation function $\chi(\omega)$ at $T=0$ for $N=10$.
 The solid line represents Eq.(\ref{chiwa}) with $\Gamma/J=3$.}
\label{t0}
\end{center}
\end{figure}

\begin{figure}[htb]
\begin{center}
\includegraphics[width=0.5\columnwidth]{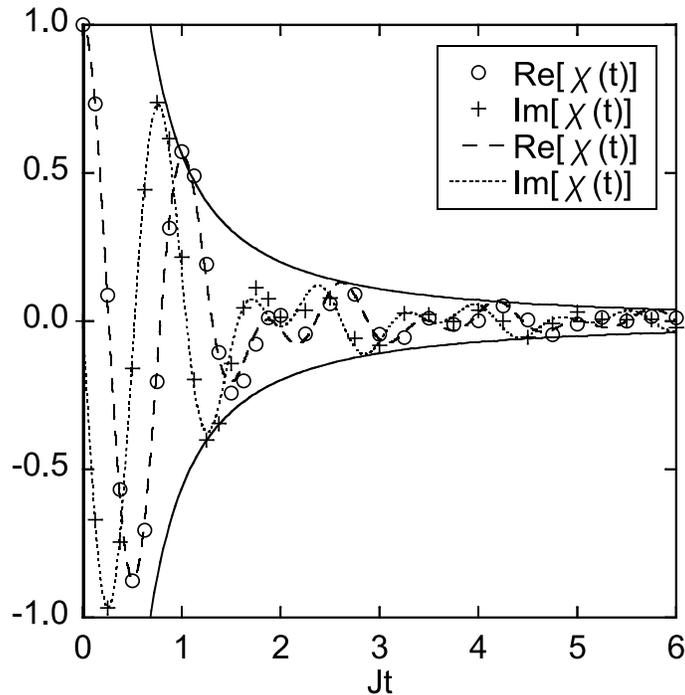}
\caption{$\chi(t)$ at $T=0$ and $\Gamma/J=3$ 
 obtained from the Fourier transformation of the data 
 in Fig.~\ref{t0}.
 The points denote numerical data and the lines are analytical results.
 The solid lines are envelope curves $\pm\pi^{-1/2} (Jt)^{-3/2}$.}
\label{ft}
\end{center}
\end{figure}

 In order to check the analytical result at $T=0$
 we carry out numerical calculation.
 The method of numerical analysis is based on the spectral decomposition
 of the correlation function 
 $\chi(\omega)=\int dt \chi(t)\exp(i\omega t)$ as 
\be
 \chi(\omega) = 
 \left[
 \frac{\sum_{nm} e^{-\beta E_n}\delta\left(\omega-(E_m-E_n)\right)
 \left|\langle n|\sigma^z_i|m\rangle\right|^2}
 {\sum_n e^{-\beta E_n}}
 \right],
 \label{sp}
\ee
 where $n$ and $m$ denote energy eigenstates. 
 This expression means that
 components in the summations contribute to $\chi(\omega)$ only
 when $\omega$ is equal to a difference of energies between two levels.
 At $T=0$, it is clear that only non-negative excitations are allowed
 and $\chi(\omega)$ forms a spectrum only in the range of positive
 $\omega$ as we showed above.

 The quantity $\chi(\omega)$ is evaluated numerically by
 the diagonalization of the Hamiltonian.
 The number of sites is taken to be $N=10$,
 and the averaging is performed over more than 10000 samples.
 The zero-temperature results are shown in Fig.~\ref{t0}
 and are compared with the analytical result (\ref{chiwa}).
 When the transverse field $\Gamma$ is not so small, 
 we find a good agreement despite the fact that Eq.(\ref{chiwa}) 
 is the Fourier transformation of the asymptotic form (\ref{chita}).
 We also note that a similar form of $\chi(\omega)$ has already 
 been obtained in larger system sizes in Ref.~\onlinecite{AR}, 
 which supports our analytical result as well.

 We also show the correlation in real-time space in Fig.~\ref{ft}
 by using numerical Fourier transformation of $\chi(\omega)$.
 The power index is estimated by linear fitting of 
 $\ln|{\rm Re}\,\chi(t)|$ and $\ln|{\rm Im}\,\chi(t)|$ peaks
 versus $\ln t$.
 Due to the finite-size effect, 
 the plot in Fig.~\ref{t0} shows small oscillations
 and smooth tails,
 which leads to large deviations at large $t$ in Fig.~\ref{ft}.
 For this reason, 
 the fitting is carried out within the region $t<5$. 
 The power index is roughly estimated as 1.6,
 which is consistent with the analytical result, 1.5.
 We expect the small deviation from the analytical one 
 also arises from the finite-size effect in small $t$ region, which
 is relatively smaller than in large $t$ region.

 As in Fig.~\ref{t0},
 when the transverse field $\Gamma$ is decreased 
 the spectral band around $\omega=2\Gamma$ approaches the origin.
 We also observe a small sharp peak in the band 
 around  a certain value of $\omega$ with $\omega<2\Gamma$.
 This peak grows up and approaches the origin 
 with increasing system size $N$. 
 We expect this peak finally reaches the origin in infinite
 system size limit.
 Since the spin-glass order parameter $q$ is equal to 
 the static part of $\chi(\tau)$ at $T=0$,~\cite{KT} 
 $\chi(\omega=0)>0$ implies the spin-glass phase.
 Therefore we can interpret the emergence of the peak
 as a sign of the phase transition from the paramagnetic phase
 to the spin-glass phase.~\cite{MH,AR}
 Although our system size $N=10$ is not large enough 
 to find the transition point, 
 the transition seems to happen at a value around
 $\Gamma\sim 1.5J$, which is consistent with the previous analyses.
 Since the main aim of the present paper is not to determine
 the precise value of the transition point, 
 we defer the detailed analysis to a future work.

\section{Correlations at finite temperature}
\label{finite}

\subsection{Numerical results}


\begin{figure}[tb]
\begin{center}
\includegraphics[width=0.5\columnwidth]{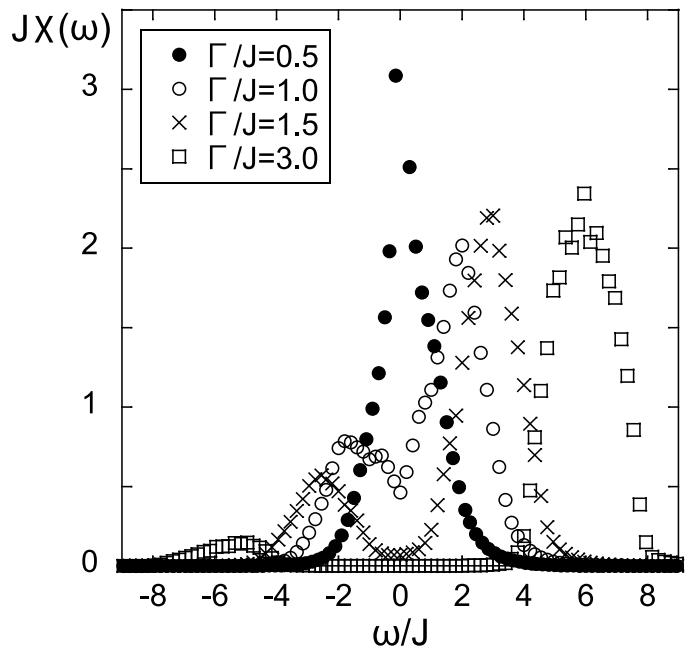}
\caption{$\chi(\omega)$ at $T=2.0J$ for $N=10$.}
\label{t2}
\end{center}
\end{figure}

\begin{figure}[htb]
\begin{center}
\includegraphics[width=0.5\columnwidth]{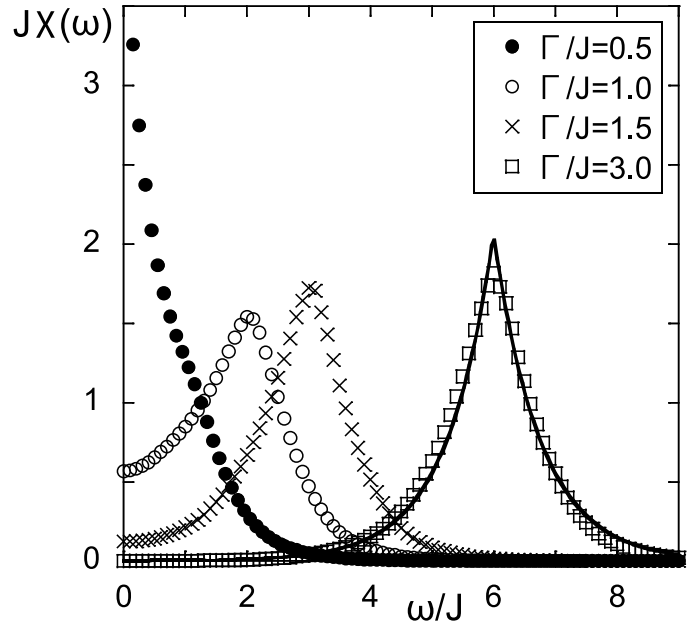}
\caption{$\chi(\omega)$ at $T=\infty$ for $N=10$.
 The result is symmetric in $\omega$ as $\chi(-\omega)=\chi(\omega)$.
 The solid line is the result of the fitting using the function in 
 Eq.(\ref{chiwinf}) with $\bar{J}= 0.76J$.}
\label{tinf}
\end{center}
\end{figure}


 Next we move on to the finite temperature case.
 When the temperature is nonzero, the negative excitations are allowed
 and we expect double peaks around $\omega=\pm 2\Gamma$.
 These peaks finally become symmetric in the limit of $T \rightarrow \infty$, 
 which can be understood from the spectral decomposition (\ref{sp}).
 The behavior at finite temperature is difficult to analyze 
 since many energy levels contribute to the correlation function.
 In the imaginary-time path-integral formalism,
 the length of the integral path shrinks as the temperature is increased
 and the assumption of asymptotic decay in $\tau$ becomes invalid.
 For these reasons we do not apply imaginary-time path-integral formalism
 to finite temperature case.

 We first observe the shape of $\chi(\omega)$
 by the numerical method instead.
 The results at $T=2.0J$ and $T=\infty$ are shown 
 in Figs.~\ref{t2} and \ref{tinf}, respectively.
 As we expect, the double peaks are formed at finite temperature.
 When $\Gamma$ is not so small,  
 the result at $T=\infty$ can be well fitted by 
 the function 
\be
 \chi(\omega) = \frac{\pi}{2\bar{J}}\left(
 e^{-|\omega-2\Gamma|/\bar{J}}
 +e^{-|\omega+2\Gamma|/\bar{J}}
 \right), \label{chiwinf}
\ee
 with a parameter $\bar{J}$.
 The result of fitting is also shown in Fig.~\ref{tinf}. 
 The parameter is taken as $\bar{J}= 0.76J$, 
 irrespective of the value of $\Gamma$.
 This form of fitting function is interpreted in the following.

\subsection{Behavior at infinite temperature}

 Let us consider a perturbative analysis 
 of the spectral representation at $T=\infty$ 
\be
 \chi(\omega) = 
 \left[\frac{1}{2^N}
 \sum_{nm} \delta\left(\omega-(E_m-E_n)\right)
 \left|\langle n|\sigma^z_i|m\rangle\right|^2
 \right]. \label{spinf}
\ee
 In the limit of $J=0$, 
 the eigenstates of the unperturbed Hamiltonian 
 $\hat{H}_0=-\Gamma\sum_{i=1}^N\sigma_i^x$
 are expressed by the eigenstates of $\sigma_i^x$.
 When the numbers of the spins pointing 
 in the negative and the positive $x$ directions 
 are $k$ and $N-k$, respectively,
 the unperturbed eigenenergy 
 is given by $E^{(0)}_k=-(N-2k)\Gamma$ and 
 the degree of degeneracy is ${}_NC_k=N!/k!(N-k)!$.
 The operator $\sigma_i^z$ flips the $i$th site spin 
 and changes the number $k$ to $k\pm 1$,
 which determines the selection rule 
 between the state $n$ and $m$ in Eq.(\ref{spinf}).
 In the first order of perturbation theory, 
 the Hamiltonian is diagonalized in a subspace 
 characterized by the quantum number $k$.
 With this quantum number we obtain the approximate expression, 
\be
 \chi(\omega) &=& \biggl[\frac{1}{2^N}\sum_{k=0}^{N-1}
 \sum_{\mu=1}^{{}_NC_k}\sum_{\nu=1}^{{}_NC_{k+1}}
 \left\{\delta\left(\omega-2\Gamma
 -(E_{k+1,\nu}^{(1)}-E_{k,\mu}^{(1)})\right)
 +\delta\left(\omega+2\Gamma
 +(E_{k+1,\nu}^{(1)}-E_{k,\mu}^{(1)})\right)\right\}
 \nonumber\\ & & \times
 \left|\langle k+1,\nu|\sigma^z_i| k,\mu\rangle\right|^2\biggr],
 \label{chiwinf1}
\ee
 where $\mu$ and $\nu$ denote the indices of degeneracy 
 and the ranges depend on $k$.
 $E^{(1)}_{k,\mu}$ are eigenvalues at the first order and
 may be expressed by linear combinations of $J_{ij}$
 as $E^{(1)}_{k,\mu}=\sum_{ij}m_{ij}^{(k,\mu)}J_{ij}$.
 Another form of $E^{(1)}_{k,\mu}$ as a function of $J_{ij}$
 will not change the conclusion here. 
 The averaging of Eq.(\ref{chiwinf1})
 with respect to $J_{ij}$ yields a Gaussian factor as 
\be
 \left[
 \delta\left(\omega-2\Gamma-\sum_{ij} m_{ij}J_{ij}\right)
 \right]
 \propto
 \exp\left\{
 -\frac{N(\omega-2\Gamma)^2}{2J^2\sum_{ij}m_{ij}^2}
 \right\}.
\ee
 The contribution of Eq.(\ref{chiwinf1})
 mainly comes from the sector with $k\sim N/2$, which
 is understood in conjunction with the factor $1/2^N$
 in Eq.(\ref{chiwinf1}). 
 For example, in the case of $k=0$
 the number of the elements in the $\mu \nu$ sum is equal to $N$
 and is negligible at $N\to\infty$ due to the multiplication by the
 factor $1/2^N$, whereas it is not suppressed when $k \sim N/2$.

 Although it is hard to perform explicitly the perturbative calculation
 even at the first order, 
 we easily see that the result is expressed by a linear combination of
 Gaussian functions as 
\be
 \chi(\omega) = 2\pi\int d\sigma P(\sigma)
 \left[
 \exp\left\{-\frac{(\omega-2\Gamma)^2}{2\sigma^2}\right\}
 +\exp\left\{-\frac{(\omega+2\Gamma)^2}{2\sigma^2}\right\}
 \right],
\ee
 where $P(\sigma)$ is the distribution function of the variance $\sigma$. 
 From this expression, we find that result (\ref{chiwinf}) can be
 reproduced by assuming 
 the Gaussian distribution $P(\sigma)\propto \exp(-\sigma^2/2\bar{J}^2)$.
 The Fourier transformation of Eq.(\ref{chiwinf}) gives 
\be
 \chi(t) = \frac{\cos(2\Gamma t)}{1+\bar{J}^2t^2}.
 \label{chitinf}
\ee
 This shows that the correlation asymptotically decays as $t^{-2}$.
 We must note that this index is obtained by a different mechanism
 as Ref.~\onlinecite{MH}
 where the same value of the index is concluded at zero temperature.

\subsection{Classical and quantum paramagnetic phase}
\label{para}

 At finite temperature, the spin-glass phase is suppressed 
 in increasing $T$ and disappears at $T=J$.~\cite{MPV,Nishimori} 
 The second-order phase-transition curve,
 namely, the spin-glass phase boundary, runs from $T=0$, $\Gamma\sim
 1.5J$ to $T=J$, $\Gamma=0$ on $T$-$\Gamma$ plane.
 On the other hand, in Sec.~\ref{zero} we saw the change 
 in the behavior of $\chi(t)$ at zero temperature when 
 $\Gamma$ is varied, and this change is expected to be persistent 
 up to higher-temperature region than the spin-glass phase boundary.
 We can attribute this change to the one
 in moving across the boundary between the quantum paramagnetic (QP)
 and the classical paramagnetic (CP) phases, rather than
 between the spin-glass phase and one of the paramagnetic phases.

 The existence of the phase transition between the CP and the QP phases
 has already been found in the $p$-body interaction spin-glass model
 in a transverse field defined by the Hamiltonian
\be
 \hat{H}=-\sum_{i_1<i_2<\cdots<i_p}J_{i_1i_2\cdots i_p}
 \sigma^z_{i_1}\sigma^z_{i_2}\cdots\sigma^z_{i_p}
 -\Gamma\sum_i\sigma^x_{i}.
\ee
 The interaction $J_{i_1i_2\cdots i_p}$ 
 is random and the average is taken with 
 Gaussian probability distribution.
 At $p\to \infty$ this model is known as the random energy model
 and exhibits first-order phase transition 
 between the CP and the QP phases.~\cite{REM}
 The imaginary-time-correlation function at the QP phase can be calculated 
 from the self-consistent equation in terms of $\chi(\tau)$, 
\be
 \chi(\tau) = \left<
 s^{(1)}_{z}(\tau)s^{(1)}_{z}(0)
 \exp\left\{
 \frac{pJ^2}{4}\sum_{\alpha=1}^n\int_0^\beta d\tau_1\int_0^\beta d\tau_2\,
 s_{z}^{(\alpha)}(\tau_1)\chi^{p-1}(\tau_1-\tau_2)s_{z}^{(\alpha)}(\tau_2)
 \right\}
 \right>_\Gamma.
\ee
 The factor $\chi^{p-1}$ goes to zero at $p\to\infty$ unless $\chi=1$. 
 From this, it is clear the zeroth-order calculation becomes exact, 
\be
 \chi(\tau) = D_\Gamma(\tau) 
 = \frac{e^{\beta\Gamma-2\Gamma\tau}
 +e^{-\beta\Gamma+2\Gamma\tau}}
 {e^{\beta\Gamma}+e^{-\beta\Gamma}}. \label{chiREM}
\ee
 The phase-transition point cannot be determined from this equation 
 for the QP phase but can be determined by the static part 
 of the correlation function for the CP phase. 
 Considering the condition of the vanishing static part, we obtain 
 a discontinuous change from $\chi(\tau)=1$ to Eq.(\ref{chiREM}),
 which also tells us the boundary between the CP and the QP phases
 reaches $T=\infty$.
 When $p$ takes a large but finite value,
 the phase-transition line is terminated 
 at a certain finite temperature value and the transition between 
 the CP and the QP phases turns to crossover.~\cite{p}

 Our numerical analysis for the SK ($p=2$) model 
 shows that the semicircle form changes to a broader one 
 at finite temperature.
 The long tail of the spectrum reaches the origin $\omega=0$, 
 just as Eq.(\ref{chiwinf}).
 We also observe that a sharp peak around the origin, 
 which was considered a precursor of 
 the spin-glass transition for finite systems, 
 is suppressed at finite temperature.
 These observations imply that 
 the spectrum change at finite temperature indicates 
 a crossover between the CP and the QP phases. 
 This picture is supported by
 the analysis in Ref.~\onlinecite{KT} where 
 a smooth change in the static part of the correlation function 
 was obtained at $T>0$.
 However, it is known in a Langevin dynamics model 
 of $p$-body interaction spin glass \cite{p2}
 that the behavior at $p>2$ should be distinguished from 
 the one with $p=2$.
 Our numerical calculation is carried out for rather small systems, 
 and it is fair to say that our analysis is not conclusive 
 and further studies are required to determine the properties of
 the finite temperature phase diagram.

\section{Conclusions}
\label{conc}

 In conclusion, we have calculated 
 the dynamical correlation function (\ref{chit})
 in the transverse SK model (\ref{GSK}).
 Our main results are Eqs.(\ref{chita}) and (\ref{chiwa}) at $T=0$,  
 and Eqs.(\ref{chiwinf}) and (\ref{chitinf}) at $T=\infty$.
 For large transverse field, the correlation function asymptotically
 decays in time $t$ as $t^{-3/2}$ at $T=0$ 
 and $t^{-2}$ at $T=\infty$.
 The results of our analysis are different
 from those of previous works at $T=0$.~\cite{MH,YSR,RG,GR} 
 In the present spin system,
 it is the crucial point in the analysis
 that the simple Wick theorem cannot be applied to
 the multicorrelations of the spin operators. 
 It was shown in Ref.~\onlinecite{WSC} that the factorization for 
 Gaussian variables is modified in the case of spin operators
 and a sign factor is introduced to respect time ordering
 of operators.
 We showed that the correct procedure leads to a different form 
 of real-time correlation from the known one.

 In the present paper, we have analyzed the infinite range model 
 to focus on the time dependence of the correlation function.
 It is important to study how the present result is changed 
 at finite-dimensional models with finite range interactions.
 Spatial correlation can be related to time correlation; 
 therefore the result of such study will be useful.
 With regard to critical exponent, there is an argument that
 the dynamic critical exponent $z$ is related to the power
 index of the time-correlation function
 discussed in the present paper.~\cite{ARi} 
 In addition, in low-dimensional systems 
 it is inferred from the general considerations \cite{Sachdev}
 that correlation function is
 significantly affected by Griffiths-McCoy singularities.~\cite{GM}

 Application to other random quantum systems  
 is also an interesting future work.
 For the random quantum Heisenberg model, the correlation 
 function in Fourier-transformed space, $\chi(\omega)$, was 
 numerically analyzed in Refs.~\onlinecite{AR} and \onlinecite{AR2}.
 It is interesting to analyze the correlation in real-time space 
 by using the analytical method developed here, 
 and this will be our future work.

\section*{ACKNOWLEDGMENTS}

 We are grateful to 
 H. Nishimori and T. Obuchi for useful discussions and comments.
 One of the authors (K. Takeda) is supported by Grand-in-aid 
 from MEXT/JSPS, Japan (Grant No.18079006).


\end{document}